\documentclass[aps,prl,twocolumn,superscriptaddress,showpacs]{revtex4-1}
\usepackage{amssymb,amsmath,amsfonts,graphicx,epsf}
\usepackage{color}

\begin{document}

\title{Temporal Intermittency of Energy Dissipation in Magnetohydrodynamic Turbulence}

\author{Vladimir Zhdankin}
\email{zhdankin@wisc.edu}
\affiliation{Department of Physics, University of Wisconsin, 1150 University Avenue, Madison, Wisconsin 53706, USA}
\author{Dmitri A. Uzdensky}
\affiliation{Center for Integrated Plasma Studies, Physics Department, UCB-390, University of Colorado, Boulder, CO 80309} 
\author{Stanislav Boldyrev}
\affiliation{Department of Physics, University of Wisconsin, 1150 University Avenue, Madison, Wisconsin 53706, USA}

\date{\today}

\begin{abstract}
Energy dissipation in magnetohydrodynamic (MHD) turbulence is known to be highly intermittent in space, being concentrated in sheet-like coherent structures.  Much less is known about intermittency in time, another fundamental aspect of turbulence which has great importance for observations of solar flares and other space/astrophysical phenomena. In this Letter, we investigate the temporal intermittency of energy dissipation in numerical simulations of MHD turbulence. We consider four-dimensional spatiotemporal structures, ``flare events'', responsible for a large fraction of the energy dissipation. We find that although the flare events are often highly complex, they exhibit robust power-law distributions and scaling relations. We find that the probability distribution of dissipated energy has a power law index close to $\alpha \approx 1.75$, similar to observations of solar flares, indicating that intense dissipative events dominate the heating of the system. We also discuss the temporal asymmetry of flare events as a signature of the turbulent cascade. \\
\end{abstract}

\pacs{52.35.Ra, 95.30.Qd, 96.60.Iv, 52.30.Cv}

\maketitle

{\em Introduction.}--- 
Intermittency plays a major role in turbulence by causing processes such as energy dissipation and particle acceleration to be highly localized in coherent structures. It also forestalls efforts toward a complete theory of turbulence. Many tools have been employed to study intermittency, including structure functions \cite{muller_etal_2003, chen_etal_2011}, scale-dependent kurtosis \cite{wan_etal_2012}, topological methods \cite{servidio_etal2009, servidio_etal2010}, and statistics of discontinuities \cite{greco_etal_2009b, zhdankin_etal_2012}. However, past studies have mainly focused on spatial intermittency, giving limited information about the dynamics.  In order to understand the temporal aspects of intermittency, including characteristic timescales of structures as well as their interactions and stability, a broader framework is needed.

A promising new paradigm is the statistical analysis of coherent structures, which is robust and informative for studies of intermittency. The occurrence rates, intensities, and morphology of structures yield insight to the inhomogeneity, anisotropy, and characteristic scales of the dynamics. Coherent structures can be simply identified as regions in space bounded by an isosurface of some field. This was used to study vorticity filaments in hydrodynamic turbulence \cite{jimenez_etal1993, moisy_jimenez2004, leung_etal2012}, magnetic structures in the kinematic dynamo \cite{wilkin_etal2007}, and dissipative structures in magnetohydrodynamic (MHD) turbulence \cite{uritsky_etal2010, zhdankin_etal2013, zhdankin_etal2014, wan_etal_2014} and ambipolar diffusion MHD \cite{momferratos_etal_2014}. Since coherent structures and intense dissipative events are experimentally observable, there are many practical applications including solar flares, instabilities in fusion devices \cite{carbone_etal2000}, and radiative signatures in optically thin astrophysical plasmas, e.g., in black-hole accretion disk coronae \cite{dimatteo_etal_1999}, hot accretion flows \cite{eckart_etal_2009}, and jets \cite{albert_etal_2007};  in pulsar wind nebulae \cite[e.g.][]{tavani_etal_2011,abdo_etal_2011}; and possibly in the hot gas in galaxy clusters.

This Letter addresses some fundamental aspects of intermittency in MHD turbulence. A major question is whether, in the limit of large Reynolds number, energy dissipation is dominated by a few intense, large-scale events or by many weak, small-scale events. A related question is whether there is an inherent relationship between spatial intermittency and temporal intermittency, e.g., whether larger structures retain their coherency in time. These temporal aspects of intermittency have been practically unexplored in previous MHD studies.

In this Letter, we extend a framework previously developed for the statistical analysis of dissipative structures \cite{zhdankin_etal2014} into the temporal realm, thereby considering 4D spatiotemporal objects representing flare events. We apply this novel methodology to study intermittency in numerical simulations of strong incompressible MHD turbulence. We describe the distributions, scalings, and evolution of flare events by characterizing their length scales, durations, dissipated energies, and peak energy dissipation rates. These are the first results on the fundamental properties of the combined spatial and temporal intermittency of energy dissipation in 3D MHD turbulence.

The primary questions addressed here for MHD turbulence are also fundamental for the solar corona. In fact, our approach has strong similarities with observational studies of solar flares \cite{crosby_etal1993, shimizu1995, boffetta_etal1999, parnell_jupp2000, hannah_etal2008, christe_etal_2008, aschwanden_etal_2000b, uritsky_etal_2007, uritsky_etal2013, veronig_etal_2002} and stellar flares \cite{audard_etal_1999, pallavicini_etal_1990, gudel_etal_2003, telleschi_etal_2005}, which use the time-series of X-ray and extreme UV emissions to measure the duration, peak intensity, and fluence of flares, from which dissipated energy is inferred. For the solar corona, a measurement of central importance is the probability distribution for dissipated energy, due to its role in assessing the nanoflare model for coronal heating \cite{parker_1988, hudson_1991}.  This distribution exhibits a power law over eight orders of magnitude, with an index near $-1.8$, somewhat shallower than the critical index of $-2$ required for nanoflares to dominate the overall heating \footnote{Given the distribution for energy dissipation, $P(E) \sim E^{-\alpha}$, the critical index is derived by noting that the total energy dissipation, $E_\text{tot} \propto \int_{E_\text{min}}^{E_\text{max}} E P(E) dE$, scales with the lower bound $E_\text{min}$ if $\alpha > 2$ and with the upper bound $E_\text{max}$ if $\alpha < 2$.}.

We compare our results with the observed statistical properties of solar flares. We stress that there are several basic differences between incompressible MHD turbulence and the solar corona. In contrast to volumetrically-driven turbulence, the solar corona is modeled by force-free MHD with slowly-driven, line-tied boundaries. Furthermore, kinetic effects may become important during reconnection. Despite these differences between our simulations and the solar corona, we find that the statistical properties of flare events have multiple similarities in both cases. This suggests that MHD turbulence may play a role in the energetics of the corona \cite{georgoulis_2005, uritsky_etal_2007}, a possibility that should be investigated more carefully in future studies.

{~}\\
{\em Method.}--- 
We perform simulations of reduced MHD, applicable since the uniform background magnetic field ${\bf B}_0 = B_0 \hat z$ is strong relative to turbulent fluctuations, $B_0 \approx 5 b_\text{rms}$. The equations are \cite{biskamp2003}
\begin{eqnarray}
\left(\frac{\partial}{\partial t}\mp\boldsymbol{V}_A\cdot\nabla_\parallel +  \boldsymbol{z}^\mp\cdot\nabla_\perp\right)\boldsymbol{z}^\pm &=& -\nabla_\perp P 
+\nu\nabla_\perp^2\boldsymbol{z}^\pm +\boldsymbol{f}_\perp^\pm \nonumber \\
\nabla_\perp \cdot \boldsymbol{z}^\pm &=& 0 \label{rmhd-elsasser}
\end{eqnarray}  
where $\boldsymbol{z}^\pm=\boldsymbol{v}\pm\boldsymbol{b}$ are the Els\"asser variables (perpendicular to ${\bf B}_0$), $\boldsymbol{v}$ is the velocity field, $\boldsymbol{b}$ is the fluctuating magnetic field (in units of the Alfv\'en velocity, $\boldsymbol{V}_A={\boldsymbol{B}}_0/\sqrt{4\pi\rho_0}$, where $\rho_0$ is the uniform plasma density), $P$ is the total pressure, and $\boldsymbol{f}^\pm_\perp$ is the external forcing. We use uniform fluid viscosity $\nu$ and magnetic diffusivity $\eta$ with $\nu = \eta$. We consider structures in the current density $j = j_z = \hat{z} \cdot \nabla_\perp \times \boldsymbol{b}$; the resistive energy dissipation rate per unit volume is $\eta j^2$.

Equations (\ref{rmhd-elsasser}) are solved using a fully dealiased 3D pseudo-spectral algorithm (for details, see \cite{perez_etal2012}).  The periodic box is elongated in $\hat{z}$ by a factor of $L_\parallel/L_\perp = 6$, where $L_\perp = 2\pi$ is the box size in simulation units. Timescales are in units of eddy turnover times, $\tau_\text{eddy} = L_\perp/(2 \pi v_\text{rms}) \approx 1$. The turbulence is driven at large scales by colliding Alfv\'en waves, generated from statistically independent random forces $\boldsymbol{f}_\perp^\pm$ at wave-numbers $2\pi/L_{\perp} \leq k_{x,y} \leq 2 (2\pi/L_{\perp})$, $k_z = 2\pi/L_\|$. The forcing is solenoidal, has random Fourier coefficients taking Gaussian values that are refreshed independently approximately $10$ times per eddy turnover time, and has amplitude such that $b_\text{rms}\sim v_\text{rms}\sim 1$. The Reynolds number is given by $Re = v_{\rm rms} (L_\perp/2\pi) / \nu$.  

We analyze snapshots dumped at a cadence $(\Delta t)^{-1}$ from four simulations shown in Table~\ref{table-sims}. The main results are from runs with $512^3$ resolution, with the $Re = 1250$ case having the highest cadence and longest time interval. Due to computational constraints, $\Delta{t}$ is larger than the internal time step in the simulation. The minimum cadence required to properly track structures is estimated by requiring that the distance advected by the flow during $\Delta{t}$ is less than the typical current sheet thickness, giving $v_\text{rms} \Delta{t} < b_\text{rms}/j_\text{thr}$. The cadences are comparable to this value and the results show convergence with cadence.

\begin{table}[h!b!p!]
\caption{Simulations \newline} \label{sims}
\begin{tabular}{|c|c|c|c|c|} 
	\hline
\hspace{0.5 mm} Sim. \hspace{0.5 mm}  & \hspace{2 mm} Res. \hspace{2 mm}   & \hspace{4 mm}$Re$\hspace{4 mm}  &   \hspace{2 mm} $\Delta{t}^{-1}$ \hspace{2 mm} &   \hspace{2 mm} Time interval \hspace{2 mm} \\
	\hline
$1$ & $256^3$ & 800 & $64$ & 10.0  \\
$2$ & $512^3$ & 800 & $32$ & 12.2  \\
$3$ & $512^3$ & 1250 & $64$ & 15.6 \\
$4$ & $512^3$ & 1800 & $32$ & 12.2 \\
	\hline
\end{tabular}
\centering
\label{table-sims}
\end{table}

We refer to spatial dissipative structures in a given time snapshot as \it states\rm; they are identified as spatially-connected sets of points with current densities $|j(\boldsymbol{x})|$ exceeding a fixed threshold, $j_\text{thr}$ \footnote{Two points are considered spatially connected if separated by fewer than two grid spacings.}. Typical states are thin, ribbon-like current sheets aligned with the $z$ direction. They occupy a small fraction of volume but account for a large fraction of the overall resistive energy dissipation. Their lengths and widths span the inertial range, while thicknesses are localized inside the dissipation range \cite{zhdankin_etal2013, zhdankin_etal2014}.

Our present work extends this procedure into the temporal realm by applying a similar threshold criterion to the 4D space-time field $j(\boldsymbol{x},t)$. We refer to the resulting 4D spatiotemporal structures as {\it processes} or flare events. 
Our numerical algorithm first identifies the states in each snapshot as described above. It then connects the states in time by finding, for each given state, any other states in the two adjacent (past and future) snapshots having points connected to the same spatial region. In general, the states in one snapshot are not in bijective (i.e., one-to-one) correspondence with the states in the adjacent snapshot. This is due to {\it interactions} between structures, including mergers and divisions, and also the spontaneous {\it formation} of new states and {\it destruction} of old states. We refer to a sequence of states representing bijective evolution of a structure, beginning and ending with interactions, formation or destruction, as a {\it path}. Processes are then obtained as sets of paths connected via interactions. The natural and conservative approach for a temporal analysis is to study processes rather than individual paths, which become ambiguous upon interacting.

We characterize processes by the following quantities. The process duration $\tau$ is the time between the final state of the process and its initial state (normalized to $\tau_\text{eddy}$). The length~$L$ for a state is the maximum distance between any two constituent points (normalized to the perpendicular box size $L_\perp$). This is generalized for a process as the maximum length, $L_\text{max}$, among constituent states. The instantaneous Ohmic energy dissipation rate for a state is ${\cal E} = \int dV \eta j^2$, with integration over the constituent points of the state (normalized to average total energy dissipation rate, ${\cal E}_\text{tot} \approx 1$). This is generalized to the total dissipated energy of a process, $E = \int dt \int dV \eta j^2$, by combining the energy dissipation rates of all constituent states (normalized to $E_\text{tot} \approx {\cal E}_\text{tot}\tau_\text{eddy} \approx 1$). We also consider the peak energy dissipation rate, ${\cal E}_\text{max}$, which is the maximum energy dissipation rate among constituent states.

\begin{figure*}
\includegraphics[width=12cm]{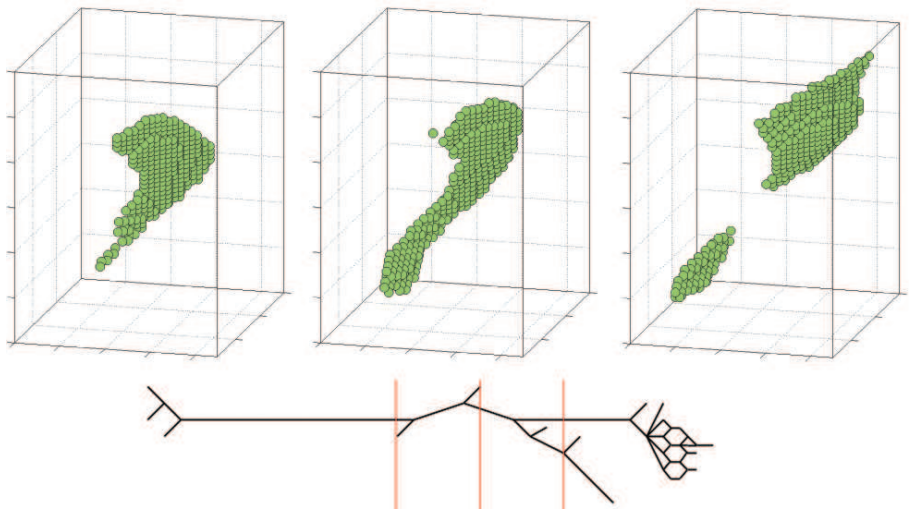}
   \centering
   \caption{\label{fig:tree} States in a typical process with duration $\tau \approx 0.5$, shown in green on a piece of the simulation lattice (with dimensions $0.10 L_\perp \times 0.14 L_\perp \times 0.90 L_\perp$, without accounting for elongation of the lattice vertically). Also shown is a schematic of paths and interactions in the process, with red lines marking the times corresponding to the shown states.}
 \end{figure*}

{~}\\
{\em Results.}--- 
For the following analysis, we choose $j_\text{thr}/j_\text{rms} \approx 6.8$. This threshold is high enough to avoid percolation of processes through space and time. Processes that exist during the initial or final snapshots do not significantly contribute to the statistical results; we retain them in distributions for better statistics.

An example process, with duration $\tau \approx 0.5$ and $31$ distinct paths, is shown in Fig.~\ref{fig:tree}. Representative states are shown (in green) on a subdomain of the simulation grid. We also show a schematic of the paths and interactions in the process. The process includes a division after the structure is stretched. A large number of paths are produced during the final stages, as the process decays toward the threshold.

We now consider the statistical properties of the processes from the four simulations in Table~\ref{table-sims}.  The mean number of states per snapshot is $\langle N_\text{state} \rangle = \{194, 288, 657, 1328\}$. For fixed cadence of $\Delta{t}^{-1} = 32$, the mean number of processes per eddy turnover time is $N_\text{proc} = \{914, 1271, 4272, 11608\}$, strongly increasing with~$Re$. The most complex processes have $\sim 10^3$ constituent paths. We find a consistent asymmetry in the interactions: there are more divisions than mergers, with a ratio $N_\text{mer}/N_\text{div} = \{ 0.84, 0.78, 0.80, 0.82\}$.

\begin{figure*}
\includegraphics[width=18cm]{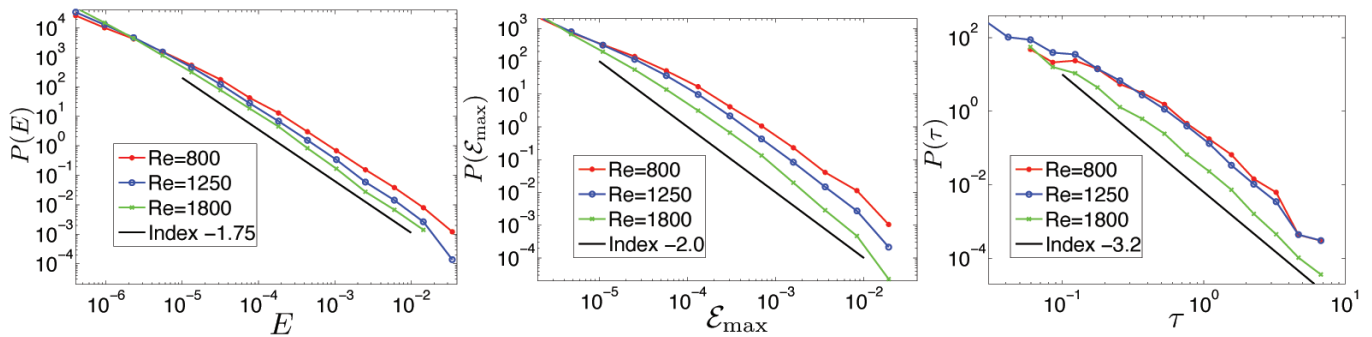}
   \centering
   \caption{\label{fig:dist_energy} The probability distributions for dissipated energy $E$, peak energy dissipation rate ${\cal E}_\text{max}$, and duration $\tau$ for $Re = 800$ (red), $Re = 1250$ (blue), and $Re = 1800$ (green).}
 \end{figure*}

We show in Fig.~\ref{fig:dist_energy} the probability distributions for dissipated energy $E$ and for peak energy dissipation rate~${\cal E}_\text{max}$. The distribution for dissipated energy, $P(E)$, has a power-law tail with an index near $-1.75\pm0.1$, which is close to the analogous observations for total energy released in solar flares \cite{aschwanden_etal_2000b, bromund_etal_1995}. The power law extends across three orders of magnitude in $E$, from $E \approx 10^{-5}$ up to about $E \approx 10^{-2}$. For smaller $E$, the distribution is shallower and apparently non-universal, likely due to dissipation-range effects and threshold effects. With increasing $Re$, the power law extends to smaller $E$, consistent with the longer inertial range. The distribution for peak energy dissipation rate, $P({\cal E}_\text{max})$, exhibits a power law with index close to~$-2.0 \pm 0.1$ from ${\cal E}_\text{max} \approx 10^{-4}$ to ${\cal E}_\text{max} \approx 10^{-2}$. Similar indices are observed in distributions for peak hard X-ray flux in solar flares \cite[e.g.,][]{bromund_etal_1995} and for energy dissipation rates ${\cal E}$ of states \cite{zhdankin_etal2014}.

The distribution for process durations $\tau$ is also shown in Fig.~\ref{fig:dist_energy}. The durations extend to well above an eddy turnover time, sometimes comparable to the analyzed time interval. The distribution from $\tau \approx 0.2$ to $\tau \approx 8$ can be fit to a power law with index near $-3.2 \pm 0.2$, somewhat steeper than the indices ranging between $-2.2$ and $-3.0$ for solar flare durations \cite{crosby_etal1993, bromund_etal_1995, veronig_etal_2002}, although close to the index $-3.4$ for rise times \cite{christe_etal_2008}.

\begin{figure*}
\includegraphics[width=16cm]{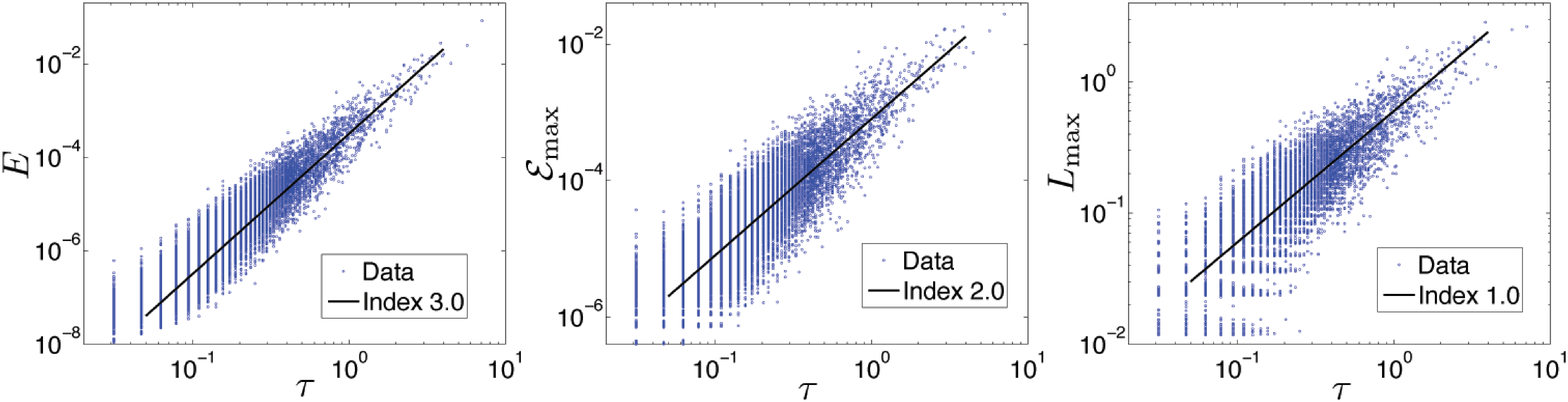}
   \centering
   \caption{\label{fig:scales_vs_duration} Scatterplots of maximum length $L_\text{max}$, peak energy dissipation rate ${\cal E}_\text{max}$, and dissipated energy $E$ versus the process duration $\tau$ for $Re=1250$.}
 \end{figure*}

The process characteristics are related by strong correlations, with examples shown in the scatterplots in Fig.~\ref{fig:scales_vs_duration}. We find that $L_\text{max} \sim \tau$, previously inferred in solar flare observations \cite{aschwanden_etal_2008}, while ${\cal E}_\text{max} \sim \tau^2$ and $E \sim \tau^3$. These scalings are consistent with the estimate $E = \int dt \int dV \eta j^2 \sim \tau {\cal E}_\text{max} \sim \tau V_\text{max} \eta j_\text{thr}^2 \sim \tau^3$, assuming that volume scales as length squared \cite{zhdankin_etal2013, zhdankin_etal2014} and current densities are near $j_\text{thr}$.  From these correlations, $E \sim {\cal E}_\text{max}^{3/2}$, and hence $P(E)$ is shallower than $P({\cal E}_\text{max})$.

\begin{figure}
\includegraphics[width=8cm]{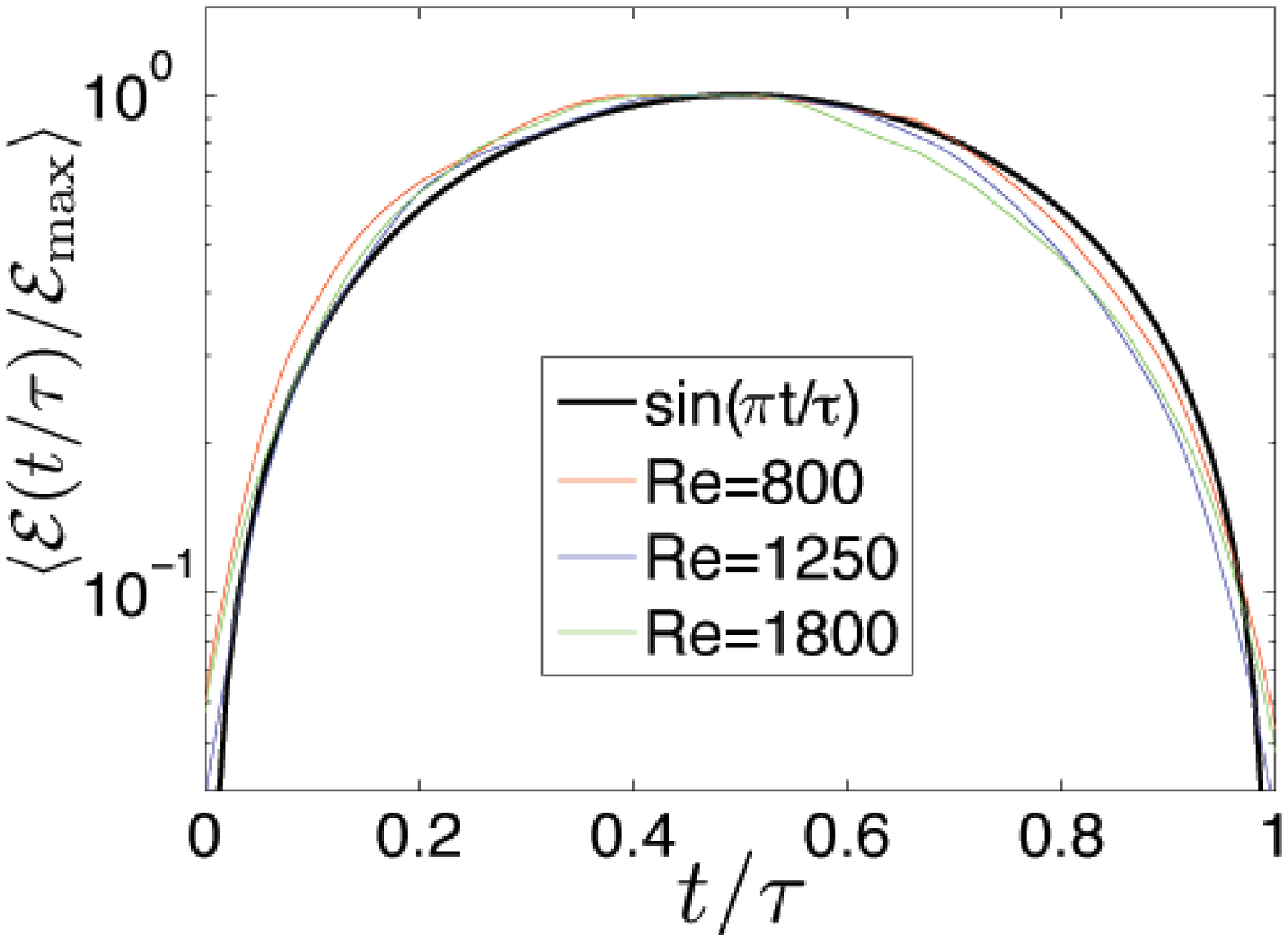}
   \centering
   \caption{\label{fig:history_sample} The evolution of energy dissipation rate relative to the peak, ${\cal E}(t/\tau)/{\cal E}_\text{max}$, averaged across all processes of all durations, for $Re = 800$ (red), $Re = 1250$ (blue), and $Re = 1800$ (green). This is well fit by a sine function (black), with small asymmetric deviations.}
 \end{figure}

Finally, we consider the evolution of processes over their durations, based on the constituent states at the given times. We focus on the energy dissipation rate, ${\cal E}(t)$ for $0 < t < \tau$. Although this time-series is irregular for any given process, we find that processes of all durations exhibit similar average evolutions. In Fig.~\ref{fig:history_sample}, we show the energy dissipation rate normalized to the peak, ${\cal E}(t)/{\cal E}_\text{max}$, versus normalized time, $t/\tau$, averaged for all processes. We find that $\langle{\cal E}(t/\tau)/{\cal E}_\text{max}\rangle \approx \sin{(\pi t/\tau)}$. Minor deviations suggest a temporal asymmetry, quantifiable by the first moment, $\langle t/\tau \rangle_{\cal E} = \int_0^\tau (t/\tau) {\cal E}(t) dt / \int_0^\tau {\cal E}(t) dt$,
which is $0.5$ for symmetric functions. We measure a very small but consistent asymmetry: for $Re = \{ 800,1250,1800 \}$, $\langle t/\tau \rangle_{\cal E} = \{0.483,0.483,0.476\}$. Hence a flare event grows slightly faster than it decays; this is similar to observed solar and stellar flares, although the asymmetry is amplified by the Neupert effect \cite{aschwanden_etal_2000b, telleschi_etal_2005, temmer_etal_2001, veronig_etal_2002, christe_etal_2008}.

{~}\\
{\em Conclusions.}---
In this Letter, we investigate the combined temporal and spatial intermittency of energy dissipation in numerical simulations of MHD turbulence. The conclusions are robust with respect to cadence, resolution, and threshold. We find that a significant fraction of energy dissipation occurs in current sheets that are temporally organized in intense, long-lived flare events with durations that may span several large eddy turnover times. The process duration scales proportionally to its maximum length. The energy dissipated in these intense processes is distributed as a power law with index near~$-1.75$, implying the dominance of large, intense flare events. This can be compared to the spatial structures at fixed times, which have a distribution of energy dissipation \it rates \rm closer to the critical index of~$-2$, suggesting that structures of all intensities instantaneously contribute equally to the overall energy dissipation rate \cite{zhdankin_etal2014}.

We find that the distributions and scalings are insensitive to $Re$, suggesting universality. This is consistent with the fact that the structures are coherent across inertial-range scales, with only their thickness being set by the dissipation mechanism. The dependence of the statistics on $Re$ has likely saturated at the relatively low $Re$ considered here, making the results relevant for large $Re$ in space and astrophysical turbulence.

We find asymmetry in the divisions and mergers of current sheets, as well as in the evolution of the energy dissipation rate of a process. This temporal asymmetry can possibly be linked to the direct cascade of energy from large to small scales, giving large structures a tendency to divide into smaller structures. In future studies, temporal asymmetry may be a useful diagnostic for current sheet instabilities, including the tearing instability  \cite{bhattacharjee2009, uzdensky2010, huang_bhattacharjee_2012, loureiro2012}.

The present work lays out the foundation for a comprehensive statistical analysis of dissipative processes in MHD turbulence, to appear in a future paper \cite{zhdankin_etal_inprep}. The methodology can be applied to many other systems, including hydrodynamic turbulence \cite{mcwilliams_etal1999}, line-tied MHD \cite{rappazzo_velli_2011, wan_etal_2014}, kinetic plasma turbulence \cite{leonardis_etal_2013, karimabadi_etal_2013}, avalanching systems \cite{bregman_gedalin_2008}, and other complex dynamical systems.

\acknowledgements

The authors would like to thank Jean Carlos Perez for his support in conducting the numerical simulations. This work was supported by NASA grant No. NNX11AE12G, US DOE award DESC0003888, and the NSF Center for Magnetic Self-Organization in Laboratory and Astrophysical Plasmas at U. Wisconsin-Madison.


%

\end{document}